\documentclass[ALICE,manyauthors]{cernphprep}

\newcommand{\pt}{$p_{\mathrm t}$}
\newcommand{\GeV}{GeV/$c$}
\newcommand{\RAA}{$R_{\rm{AA}}$}
\newcommand{\KS}{$\rm K^0_{\rm{S}}$}
\usepackage{hyperref}
\usepackage{cite}
\usepackage{rotating}

\begin{document}%
%
%
\begin{titlepage}
\PHnumber{2012-155}      
\PHdate{25 July 2012}              
%
%
\title{Production of muons from heavy flavor decays at forward rapidity in pp 
and Pb--Pb collisions at $\sqrt {s_{\rm NN}}$~=~2.76~TeV}
\ShortTitle{Heavy flavor decay muons in pp and Pb--Pb 
at $\sqrt {s_{\rm NN}}$~=~2.76~TeV}  
%
\Collaboration{ALICE Collaboration%
         \thanks{See Appendix~\ref{app:collab} for the list of collaboration 
                      members}}
\ShortAuthor{ALICE Collaboration}  
%
\begin{abstract}
The ALICE Collaboration has measured the inclusive production of muons from 
heavy flavor decays at forward rapidity, $2.5 < y <4$, in pp and Pb--Pb 
collisions at $\sqrt {s_{\rm NN}}$ = 2.76 TeV. The \pt-differential inclusive 
cross section of muons from heavy flavor decays in pp collisions is compared 
to perturbative QCD calculations. The nuclear modification factor is studied 
as a function of \pt\ and collision centrality. A weak suppression is measured 
in peripheral collisions. In the most central collisions, a suppression of a 
factor of about 3--4 is observed in $6 < p_{\rm t} < 10$~\GeV. The suppression 
shows no significant \pt\ dependence. 
%
\end{abstract}
\end{titlepage}
\setcounter{page}{2}
The study of ultra-relativistic heavy ion collisions is aimed at investigating 
the properties of strongly-interac\-ting matter in the extreme conditions of 
high temperature and energy density expected to be reached. Quantum 
Chromodynamics (QCD) calculations on the lattice predict under such 
conditions the formation of a deconfined partonic phase, the Quark-Gluon 
Plasma (QGP), and chiral symmetry is restored~\cite{kar06}. Heavy quarks 
(charm and beauty), abundantly produced at the Large Hadron Collider (LHC), 
are sensitive probes of the properties of the QGP. Due to their large masses, 
they are created mainly in hard scattering processes during the early stage of 
the collision and subsequently interact with the hot and dense medium. 
In particular, measurement of open heavy flavor hadrons may probe the energy 
density of the system through the mechanism of in-medium energy loss of heavy 
quarks. The in-medium effects are usually quantified by means of the nuclear 
modification factor $R_{\rm AA}$ of the transverse momentum (\pt) 
distribution. Using the nuclear overlap function from the Glauber 
model~\cite{mil07}, $R_{\rm AA}$ can be expressed as:
\begin{equation}
R_{\rm AA}(p_{\rm t}) = {1\over{\langle T_{\rm AA} \rangle}} \cdot 
{{\rm d}N_{\rm AA}/{\rm d} p_{\rm t} 
\over {{\rm d}\sigma_{\rm pp}/{\rm d} p_{\rm t} }},
\label{eq:RAADef2}
\end{equation}
where $\langle T_{\rm AA} \rangle$ is the average nuclear overlap function 
in a given centrality class. The term ${\rm d}N_{\rm AA}/{\rm d} p_{\rm t}$ 
is the \pt-differential yield in nucleus-nucleus (AA) collisions, while 
${\rm d}\sigma_{\rm pp}/{\rm d} p_{\rm t}$ is the \pt-differential inclusive 
cross section in pp collisions. The value of $R_{\rm AA}$ is unity for 
hard probes if no nuclear modification is present. A $R_{\rm AA}$ value 
smaller than unity can arise from partonic energy loss as well as other 
nuclear effects. According to QCD, the radiative energy loss of gluons 
should be larger than that of quarks, and due to the dead cone 
effect~\cite{dok01,djo03,zha03,arm04}, heavy quark energy loss should be 
further reduced with respect to that of light quarks. The contribution from 
other interaction mechanisms, for instance collisional 
energy loss~\cite{mus04,wic07}, in-medium fragmentation, recombination and 
coalescence~\cite{adi07,gre04,hee06}, could also lead to a modification of 
heavy-flavor hadron \pt\ distributions in AA collisions. Finally, initial 
state effects~\cite{kha03,arm06} could complicate the interpretation of any 
deviation from unity of the $R_{\rm AA}$ in terms of energy loss effects, 
particularly in the low \pt\ region. The study of p--A collisions is required 
to quantify the role of initial state effects. 
The PHENIX and STAR Collaborations have reported a strong suppression of 
electrons from heavy flavor decays at mid-rapidity, in central Au--Au 
collisions at $\sqrt {s_{\rm NN}}$ = 200~GeV at 
RHIC~\cite{phe0,phe1,phe2,star}. 
The PHENIX Collaboration also measured a significant suppression of 
muons from heavy flavor decays at forward rapidity in 
central Cu--Cu collisions at $\sqrt {s_{\rm NN}}$ = 200~GeV~\cite{phe3}. 
Recently, a significant suppression of $\rm D$ mesons~\cite{Dmes12} and 
$\rm J/\psi$'s from $\rm B$ decays~\cite{cms} was measured at mid-rapidity 
in central Pb--Pb collisions at $\sqrt{s_{\rm NN}}$ = 2.76~TeV by 
ALICE and CMS at the LHC, respectively. A complementary measurement of heavy 
flavor suppression at forward rapidity, at the same energy, is of great 
interest in order to provide new constraints on models which aim at describing 
the nuclear modification factor as partonic energy loss. 

In this Letter, we report the first measurement at the LHC of the production 
of muons from heavy flavor decays at forward rapidity ($2.5 < y < 4$), 
with the ALICE experiment~\cite{aam08}, in pp and Pb--Pb collisions at 
$\sqrt{ s_{\rm NN}}$ = 2.76 TeV. The measured \pt-differential inclusive 
cross section of muons from heavy flavor decays in pp collisions at 
$\sqrt s$ = 2.76 TeV is compared to perturbative QCD (pQCD) calculations. 
In-medium effects are investigated by means of the nuclear modification 
factor as a function of \pt\ in $4 < p_{\rm t} < 10$ \GeV, and as a function 
of collision centrality in $6 < p_{\rm t} < 10$~\GeV.

The ALICE experiment is described in detail in~\cite{aam08}. The apparatus is 
composed of a central barrel (pseudo-rapidity coverage 
$\vert \eta \vert < 0.9$), a muon spectrometer 
($ -4 < \eta < -2.5$~\cite{note}) and a set of detectors for global collision 
characterization and triggering located in the forward and backward 
pseudo-rapidity regions. The VZERO, made of two scintillator arrays covering 
the $2.8 < \eta < 5.1$ and~$-3.7 < \eta < -1.7$, is used for triggering, 
centrality determination and background removal. The two Zero Degree 
Calorimeters (ZDC), located at $\pm$ 114~m from the 
interaction point, are used in offline rejection of background events. 
The Silicon Pixel Detector (SPD), a two-layer central barrel that constitutes 
the innermost part of the Inner Tracking System (ITS), is included in the 
trigger logic. The SPD provides also the interaction vertex reconstruction. 
The muon spectrometer consists of a 10 interaction length ($\lambda_{\rm I}$) 
passive front absorber, a beam shield, an iron wall, a 3~T$\cdot$m dipole 
magnet and a set of tracking and trigger chambers. Tracking is performed by 
means of five stations of cathode pad chambers, with the third station inside 
the dipole magnet. The tracking system is supplemented by two trigger stations 
of resistive plate chambers, behind a 1.2~m thick iron wall with thickness 
7.2 $\lambda_{\rm I}$. The latter absorbs hadrons that punch through the 
front absorber, as well as secondary hadrons produced inside it and low 
momentum muons, mainly from pion and kaon decays.

The Pb--Pb data were collected during the 2010 run. The rate of hadronic 
collisions was about 100 Hz, corresponding to a luminosity of 
$1.3 \cdot 10^{25}\ {\rm cm}^{-2} {\rm s}^{-1}$. The results presented 
in this Letter are based on the analysis of minimum bias (MB) trigger events. 
The MB trigger required the following conditions: a signal in at least two 
pixel chips in the outer layer of the SPD and a signal on each 
VZERO detector. The beam-induced background was reduced by using the timing 
information from the VZERO and ZDC detectors, and by exploiting the 
correlation between the number of hits and track segments in the SPD. 
Moreover, a minimal energy deposit in the ZDC was required in order to 
reject electromagnetic interactions. Finally, only events with an interaction 
vertex within $\pm 10$~cm from the center of the detector along the beam line 
were analyzed. Pb--Pb collisions were classified according to their degree 
of centrality by means of the sum of amplitudes of the signals in the VZERO 
detectors, as described in~\cite{cc1,cc2}. The analysis was limited to 
the 80$\%$ most central events for which the MB trigger was fully efficient. 
This leads to a data sample of $16.6 \cdot 10^6$ Pb--Pb collisions which, in 
the following, will be divided into five centrality classes: 0--10$\%$, 
10--20$\%$, 20--40$\%$, 40--60$\%$ and 60--80$\%$ (the two last bins will be 
grouped together for the study of $R_{\rm AA}$(\pt)). The corresponding 
integrated luminosity is $L_{\rm int}$ = $2.71 \pm 0.09\ \mu {\rm b}^{-1}$.
The values of the mean number of participating nucleons and mean nuclear 
overlap function are given in Table~\ref{tab:cent}. They were determined with 
the Glauber Monte-Carlo simulation assuming an inelastic nucleon-nucleon 
cross section of 64~mb~\cite{cc1}.
\begin{table}[hb!]
\caption{\label{tab:cent} Mean number of participating nucleons 
($\langle N_{\rm part} \rangle$) and mean nuclear overlap function 
($\langle T_{\rm AA} \rangle$) for different centrality classes, 
expressed in percentiles of the hadronic Pb--Pb cross section. }
\centering
\begin{tabular}{ c|c|c }

Centrality & $\langle N_{\rm part} \rangle$ & 
$\langle T_{\rm AA} \rangle$ mb$^{-1}$\\ \hline
0--10$\%$ & 357 $\pm$ 4 & 23.48 $\pm$ 0.97 \\ 
10--20$\%$ & 261 $\pm$ 4 & 14.43 $\pm$ 0.57 \\ 
20--40$\%$ & 157 $\pm$ 3 & 6.85 $\pm$ 0.28 \\ 
40--60$\%$ & 69 $\pm$ 2 & 2.00 $\pm$ 0.11 \\ 
60--80$\%$ & 23$\pm$ 1 & 0.42 $\pm$ 0.03 \\ \hline 
40--80$\%$ & 46 $\pm$ 2 & 1.20 $\pm$ 0.07\\ 
\end{tabular}
\end{table}
The strategy of cuts applied to reconstructed tracks is similar to the one 
used for pp collisions~\cite{Mupp}. Various selection cuts were used in order 
to improve the purity of the data sample. Tracks were required to be 
reconstructed in the geometrical acceptance of the muon spectrometer. A track 
candidate measured in the muon tracking chambers was then required to be 
matched with the corresponding track measured in the trigger chambers. This 
results in a very effective rejection of the hadronic background that is 
absorbed in the iron wall. Furthermore, the correlation between momentum and 
Distance of Closest Approach (distance between the extrapolated muon track and 
the interaction vertex, in the plane perpendicular to the beam direction and 
containing the vertex) was used to remove the remaining beam-induced 
background tracks that do not point to the interaction vertex, and fake tracks 
(tracks not associated to one single particle crossing the spectrometer). 
After these selections, the data sample consists of $10 \cdot 10^6$ muon 
candidates. The $R_{\rm AA}$ measurement of muons from heavy flavor decays 
will be performed at high \pt\ ($p_{\rm t} >$~4-6~\GeV) where the main 
background component consists of muons from primary pion and kaon decays. 
The Pb--Pb distributions are corrected for acceptance and for tracking and 
trigger efficiency (${\rm A} \times \epsilon$) using the procedure described 
in~\cite{Mupp}. The global ${\rm A} \times \epsilon$ is close to 80$\%$ for 
$p_{\rm t} > 4$~\GeV. The dependence of the trigger and tracking efficiency 
on the detector occupancy, which is correlated with the collision centrality, 
was evaluated by means of the embedding procedure~\cite{JpsiPbPb}. 
A decrease of the efficiency of about 4$\%$~$\pm$~1$\%$ is observed in the 
10$\%$ most central collisions. 

The \RAA\ of muons from heavy flavor decays in the forward rapidity region 
is calculated according to Eq.~(\ref{eq:RAADef2}), which can be written as:
\begin{equation}
R_{\rm AA}^{\mu^\pm\leftarrow {\rm {HF}}} (p_{\rm t}) = 
{1\over {\langle T_{\rm {AA}}\rangle}} \times  
{  {\rm d}N_{\rm {PbPb}}^{\mu^\pm}/
{\rm d}p_{\rm t} - 
{\rm d}N_{\rm {PbPb}}^{\mu^\pm\leftarrow {\rm \pi^\pm,K^\pm}}/
{\rm d}p_{\rm t}
\over 
{{\rm d}\sigma_{\rm {pp}}^{\mu^\pm\leftarrow {\rm {HF}}}/
{\rm d}p_{\rm t}}}
\label{eq:RAAFW2}
\end{equation}
where ${\rm d}N_{\rm {PbPb}}^{\mu^\pm}/{\rm d}p_{\rm t}$ 
and ${\rm d}N_{\rm {PbPb}}^{\mu^\pm \leftarrow \pi^\pm, {\rm K^\pm}}
/{\rm d}p_{\rm t}$ are the inclusive muon and charged pion and kaon decay 
muon \pt\ distributions at forward rapidity in Pb--Pb collisions, 
respectively.

The pp reference, ${\rm d}\sigma_{\rm {pp}}^{\mu^\pm\leftarrow {\rm {HF}}}/
{\rm d}p_{\rm t}$, was obtained from the analysis of muon-triggered events 
collected during a pp run at $\sqrt s$~=~2.76 TeV, in March 2011, 
with integrated luminosity of 19~nb$^{-1}$ after event selection cuts. The 
analysis technique from the event and track selection to the normalization, 
is the same as that described in~\cite{Mupp}. 
 \begin{figure}[hbt!]
\centering
{\includegraphics[width=.6\textwidth]{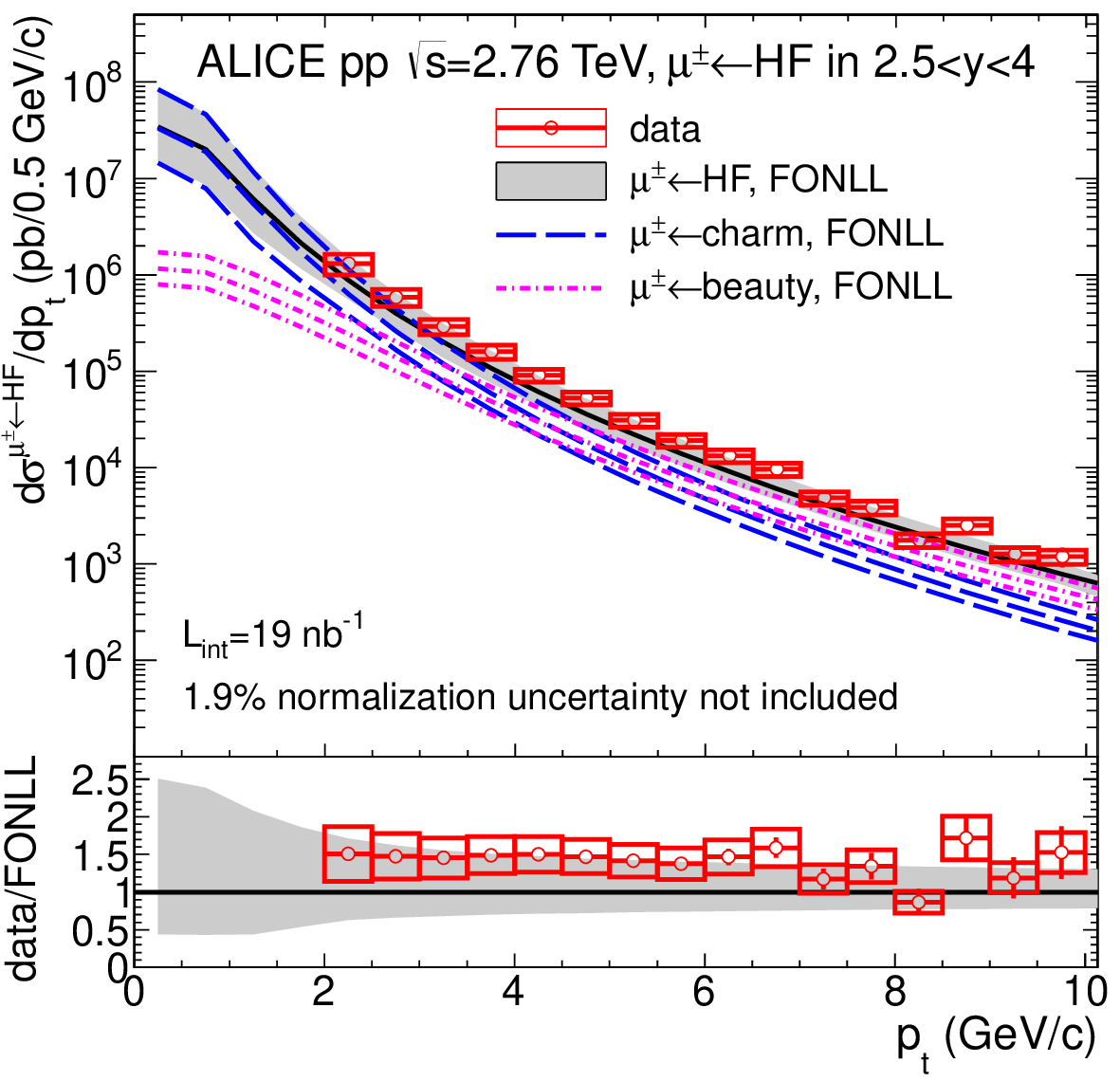}}
\caption{(Color online) Transverse momentum differential inclusive cross 
section of muons from heavy flavor decays in $2.5 < y < 4$, 
in pp collisions at $\sqrt s$~=~2.76~TeV. The vertical 
error bars (open boxes) are 
the statistical (systematic) uncertainties. The solid curve and the band show 
FONLL~\cite{fonll1,fonll2} calculations and theoretical 
uncertainties, respectively. The FONLL calculations are also reported 
for muons from charm (long dashed curves) and beauty (dot-dashed curves) 
decays, separately. The lower panel shows the ratio between data and FONLL 
calculations.}
\label{fig:ppref}
\end{figure}
Figure~\ref{fig:ppref} shows the measured~\pt-differential inclusive cross 
section of muons from heavy flavor decays in the kinematic region 
$2.5 < y < 4$ and $2 < p_{\rm t} < 10$~\GeV. In the range 
$p_{\rm t} > 4$ \GeV\ ($p_{\rm t} > 6$~\GeV), regions of interest for the 
$R_{\rm AA}^{\mu^\pm\leftarrow {\rm {HF}}} (p_{\rm t})$ measurement, the 
contribution of muons from primary light hadron decays (mainly primary pion 
and kaon decays) that was subtracted amounts to about 19$\%$ (12$\%$) of the 
total yield. The error bars are statistical uncertainties. The open boxes 
represent the systematic uncertainties varying from 15$\%$ to 24$\%$, 
depending on \pt. This includes the contributions from background subtraction 
(ranging from a maximum of about 24$\%$ at \pt\ = 2 GeV/$c$ to 14$\%$ 
at \pt\ = 10 GeV/$c$), detector response (3$\%$) and residual mis-alignment of 
tracking chambers (1$\%$ $\times$ \pt, in GeV/$c$). The systematic uncertainty 
on the minimum bias pp cross section (1.9$\%$), used in the normalization, 
is not shown. The data are compared to Fixed Order Next-to-Leading Log 
(FONLL) pQCD predictions~\cite{fonll1, fonll2} (curve, with shaded band for 
the uncertainty). The ratio between data and FONLL calculations is also shown. 
The measured \pt-differential inclusive cross section of muons from heavy 
flavor decays is well reproduced by the calculations within experimental and 
theoretical uncertainties, although at the upper limit of the predictions. 
A similar agreement between heavy flavor results and pQCD calculations 
was also reported in pp collisions at $\sqrt s$~=~7~TeV in the four 
LHC experiments and at lower energies, at the 
FNAL Tevatron and at the RHIC (see~\cite{Mupp} and references therein). 
The contributions of muons from charm and 
beauty decays from the FONLL calculations are displayed separately in 
Fig.~\ref{fig:ppref}. According to these predictions, the component of 
muons from beauty decays exceeds that of muons from charm decays 
for $p_{\rm t} \gtrsim 6$~\GeV.

The \pt\ distribution of muons from heavy flavor decays in Pb--Pb collisions 
at forward rapidity is obtained by subtracting the muon background component 
(mainly muons from primary pion and kaon decays) from the corrected inclusive 
muon \pt-differential distribution. The presence of unknown nuclear effects, 
in particular medium-induced parton energy loss at forward rapidity, prevents 
subtraction of this contribution by means of Monte-Carlo simulations, 
as was done in pp collisions~\cite{Mupp}. Hence, the contribution of muons 
from primary $\pi^\pm$ and $\rm K^\pm$ decays at forward rapidity in Pb--Pb 
collisions was estimated by extrapolating to forward rapidity ($2.5 < y <4$) 
the \pt\ distributions of pions and kaons measured at central rapidity 
($\vert y \vert < 0.8$) in pp and Pb--Pb collisions~\cite{app11} and 
generating the corresponding \pt\ distributions of decay muons with a 
simulation of the decay kinematics and of the front absorber. For the rapidity 
extrapolation, it was assumed that the suppression of pions and kaons is 
independent of rapidity up to $y = 4$. This assumption is motivated by the 
observation, by the~ATLAS Collaboration, that the central-to-peripheral 
nuclear modification factor of charged hadrons does not show any 
$\eta$-dependence up to $\eta = 2.5$ within uncertainties~\cite{atlas}. 
The systematic uncertainty introduced by this assumption was conservatively 
estimated by varying $R_{\rm AA}^{\pi^\pm,{\rm K^\pm}} (p_{\rm t})$ 
from 0 (full suppression) up to two times its value. The entire 
background-estimation procedure is detailed in the following.

The \pt\ distribution of pions and kaons at forward rapidity in Pb--Pb 
collisions in a given centrality range is expressed as: 
\begin{equation}  
{\rm d}N_{\rm {PbPb}}^{\pi^\pm, {\rm K^\pm}} 
/{\rm d}p_{\rm t}  =  
\langle T_{\rm AA} \rangle  \times
{\rm d}\sigma_{\rm {pp}}^{\pi^\pm,{\rm K^\pm}} 
/{\rm d}p_{\rm t} \times 
[R_{\rm AA}^{\pi^\pm,{\rm K^\pm}}(p_{\rm t})]_{y = 0}.
\label{eq:HadronPbPbFW}
\end{equation}
The mid-rapidity pion and kaon \pt\ distributions measured in pp collisions 
were extrapolated to forward rapidity using~\cite{phepp}:               
\begin{equation}\label{eq:extrapol}  
{\rm d^2}N_{\rm {pp}}^{\pi^\pm, {\rm K^\pm}} 
/{\rm d}p_{\rm t} {\rm d}y = 
 [{\rm d^2}N_{\rm {pp}}^{\pi^\pm, {\rm K^\pm}} 
/{\rm d}p_{\rm t} {\rm d} y]_{y = 0}               
\times {\rm exp} \Big({-y^2 \over {2 \sigma^2_{\rm y}}}\Big),                  
\end{equation}
with $\sigma_{\rm y}$ = 3.18.   
The latter is the average of the values obtained with the PYTHIA~\cite{pyt} 
and PHOJET~\cite{pho} event generators. Eq.~(\ref{eq:extrapol}) assumes that 
the shape of the \pt\ distribution is independent of $y$. However, results 
from the BRAHMS Collaboration suggest a small 
dependence at large rapidities~\cite{brahms}, but the effect is 
expected to be negligible in the analysis due to the small amount of muons 
from pion and kaon decays in the \pt\ range of interest (see below).

Then, the muon \pt\ distributions in $2.5 < y <4$ in pp and Pb--Pb collisions 
were obtained by means of fast simulations using the resultant pion and kaon 
\pt\ distributions as input. The effect of the front absorber was taken into 
account by considering only pions and kaons that decay before reaching a 
distance corresponding to one interaction length in the absorber. 

The input charged pion \pt\ distributions were measured up to \pt\ = 20 \GeV\ 
for all centrality classes used in the analysis. The kaon \pt\ distributions 
were determined only at low \pt. Therefore, the \KS\ \pt\ distributions 
measured up to 16 \GeV\ were used, considering 
that N(${\rm K^+}$) + N(${\rm K^-}$) = 2 $\cdot$ N(\KS). A further 
extrapolation up to 40 \GeV, by means of a power law fit, was needed. In 
addition, the \KS\ \pt\ distributions were measured only for the 0--5$\%$ and 
60--80$\%$ centrality classes. As a consequence, the \pt\ distributions of 
muons from pion and kaon decays at forward rapidity were determined 
only in these two centrality classes. For the other centrality classes used 
in this analysis (Table~1), 
the ${\rm d}N_{\rm {PbPb}}^{\mu^\pm \leftarrow \pi^\pm, {\rm K^\pm}} 
/{\rm d}p_{\rm t}$ distributions were obtained by scaling 
the $R_{\rm AA}^{\mu^\pm \leftarrow \pi^\pm}(p_{\rm t})$ with the double ratio 
$R_{\rm AA}^{\mu^\pm \leftarrow \pi^\pm,{\rm K^\pm}} (p_{\rm t})
/R_{\rm AA}^{\mu^\pm \leftarrow \pi^\pm}(p_{\rm t})$ which was found to be the 
same in the 0--5$\%$ and 60--80$\%$ centrality classes, within a 
maximum variation of 9$\%$ included in the systematic uncertainty.
 
This procedure allowed us to estimate 
${\rm d}N_{\rm {PbPb}}^{\mu^\pm \leftarrow \pi^\pm, {\rm K^\pm}} 
/{\rm d}p_{\rm t}$ and then to deduce the nuclear modification of muons from 
heavy flavor decays at forward rapidity according to Eq.~(\ref{eq:RAAFW2}). 
The background contribution to the muon \pt\ distribution increases with 
decreasing \pt. Hence, in order to limit the systematic uncertainty on 
its subtraction, \RAA\ was computed for \pt\ $>$~4~\GeV\ where this component 
is 7$\%$ (11$\%$) of the total muon yield in central (peripheral) collisions.

The systematic uncertainties on the $R_{\rm AA}$ of muons from heavy flavor 
decays originate from the pp reference, the corresponding Pb--Pb yields and 
the average nuclear overlap function. The systematic uncertainty on 
the pp reference, previously discussed, is about 15--17$\%$ 
for \pt\ $>$~4~\GeV. The systematic uncertainty on the yields of muons from 
heavy flavor decays in Pb--Pb includes contributions from:
\begin{itemize}
\item[-] the inclusive muon yields in Pb--Pb collisions, about 6--10$\%$, 
containing the systematic uncertainty on the detector response (3.5$\%$), 
the residual mis-alignment (1$\%$~$\times$~\pt, in \GeV) and 
the centrality dependence of the efficiency determined with the embedding 
procedure (1$\%$); 
\item[-] the yields of muons from primary pion and kaon decays in pp 
collisions at forward rapidity, about 17$\%$, due to the systematic 
uncertainty on the input mid-rapidity distributions, the extrapolation 
procedure ($\sigma_{\rm y}$ parameter) and the absorber effect (pion and kaon 
mean free path in the absorber);
\item[-] the $R_{\rm AA}^{\mu^\pm \leftarrow \pi^\pm} (p_{\rm t})$, 
about 14--17$\%$, due to the systematic uncertainty on the input mid-rapidity 
pion \pt\ distributions;  
\item[-] the $R_{\rm AA}^{\mu^\pm \leftarrow \pi^\pm,{\rm K^\pm}} (p_{\rm t})/
R_{\rm AA}^{\mu^\pm \leftarrow \pi^\pm} (p_{\rm t})$ double ratio, up to 
9$\%$ at \pt\ = 10 \GeV;
\item[-] the unknown suppression at forward rapidity for muons from 
primary pion and kaon decays. As mentioned, a conservative systematic 
uncertainty was considered by varying 
$R_{\rm AA}^{\pi^\pm,{\rm K^\pm}}(p_{\rm t})$ 
from 0 to two times its value, with the additional condition that the 
upper limit does not exceed unity.
\end{itemize}
Finally, the systematic uncertainty on the normalization includes the 1.9$\%$ 
uncertainty on the minimum bias cross section measurement in pp 
collisions and the uncertainty of 4.3$\%$ (centrality class 0--10$\%$) 
to 7.3$\%$ (centrality class 60--80$\%$) on $\langle T_{\rm AA} \rangle$.

\begin{figure}[htb!]
\centering
{\includegraphics[width=1\textwidth]{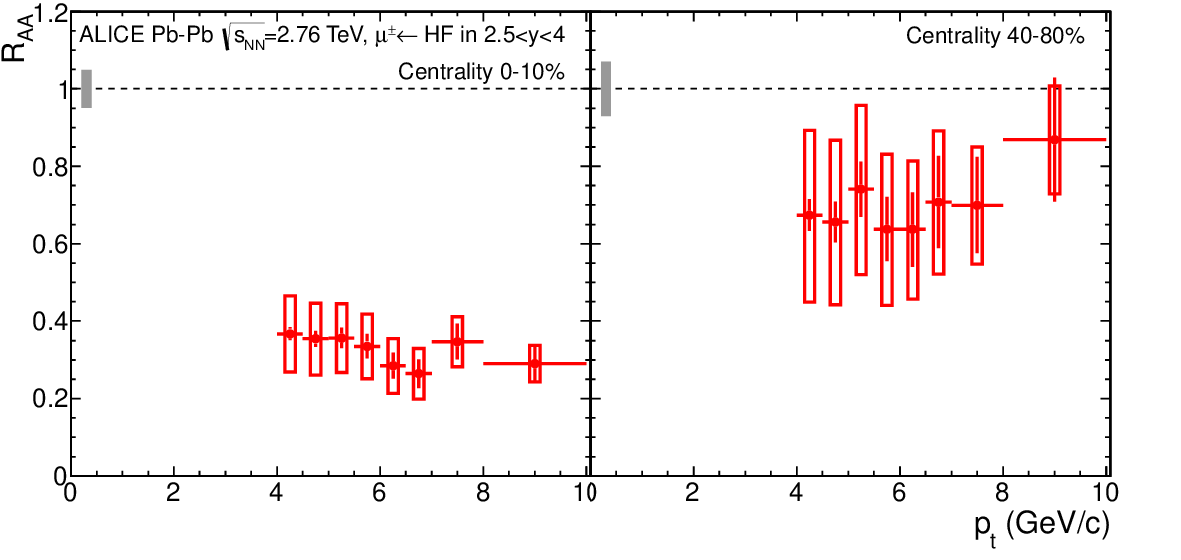}}
\caption{(Color online) \RAA\ of muons from heavy flavor decays in 
$2.5 < y < 4$ as a function of \pt, in the 0--10$\%$ (left) 
and 40--80$\%$ (right) centrality classes, in Pb--Pb collisions at 
$\sqrt {s_{\rm {NN}}}$ = 2.76 TeV. Vertical bars (open boxes) represent the 
statistical (systematic) uncertainty. The filled box centered at 
\RAA~=~1 is the normalization uncertainty. 
Horizontal bars show the bin widths.}
\label{fig:RAAMuHFPt}
\end{figure}

Figure~\ref{fig:RAAMuHFPt} presents the \RAA\ of muons from heavy flavor 
decays in $2.5 < y < 4$, as a function of \pt\ in central (0--10$\%$, left) 
and peripheral (40--80$\%$, right) collisions. The vertical error bars are 
the statistical uncertainties. The \pt-dependent systematic uncertainties 
are displayed by the open boxes and include all the contributions previously 
discussed, except the normalization uncertainty that is displayed at 
$R_{\rm AA}$~=~1. A larger suppression is observed in central collisions 
than in peripheral collisions, with no significant \pt\ dependence within 
uncertainties.

\begin{figure}[htb!]
\centering
{\includegraphics[width=.6\textwidth]{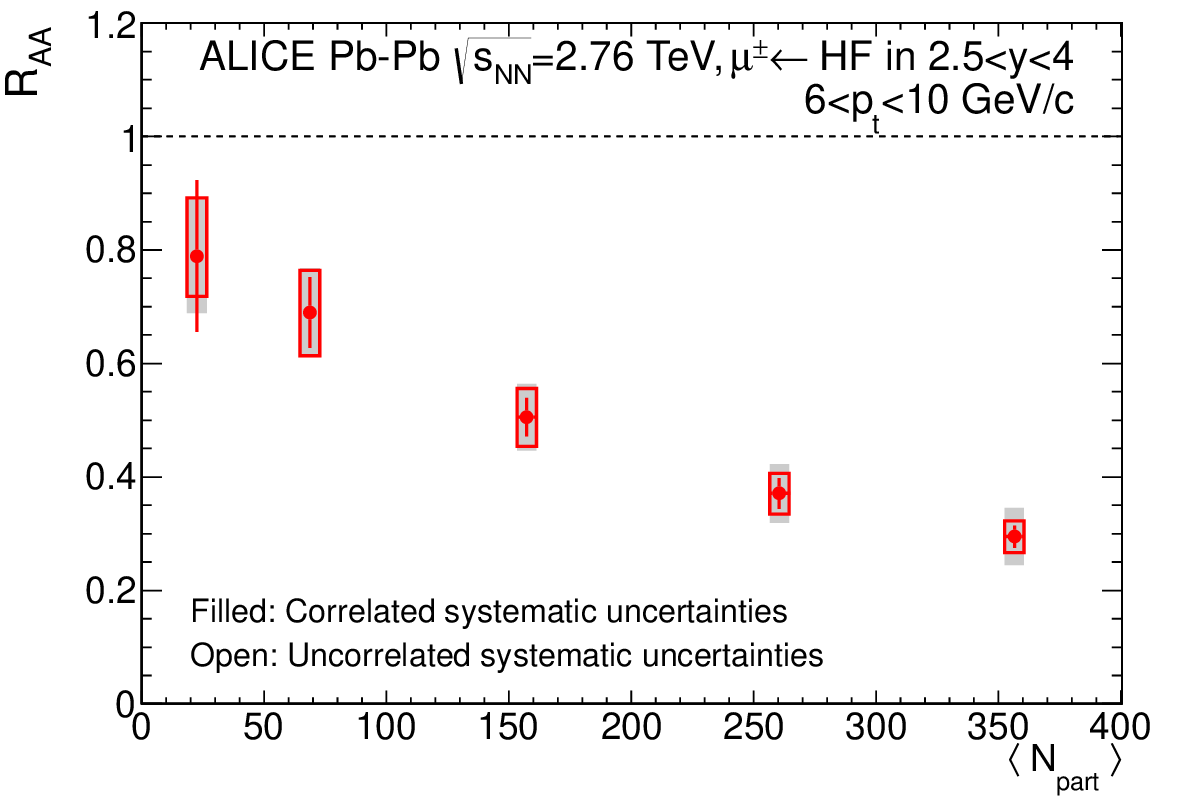}}
\caption{(Color online) \RAA\ of muons from 
heavy flavor decays as a function of the mean number of participating 
nucleons, in $2.5 <y <4$ and $ 6 < p_{\rm t} < 10$ \GeV. The horizontal bars 
indicate the uncertainty on $\langle N_{\rm part} \rangle$.} 
\label{fig:RAAMHFCC}
\end{figure}

The centrality dependence of the $R_{\rm AA}$ of muons from heavy flavor 
decays was studied in the range $6 < p_{\rm t }< 10$ \GeV\ 
where the contribution of muons from B decays becomes dominant in pp 
collisions according to the central value of the FONLL calculations: 
in particular, it amounts to about 58$\%$ and 68$\%$ at \pt\ = 6 and 
10~\GeV, respectively (Fig.~\ref{fig:ppref}). 
The analysis was carried out in five centrality 
classes from 0--10$\%$ to 60--80$\%$ (Table~\ref{tab:cent}). The resulting 
$R_{\rm AA}$ is displayed as a function of $\langle N_{\rm part} \rangle$ 
in Fig.~\ref{fig:RAAMHFCC}. The contribution to the total systematic 
uncertainty which is fully correlated between centrality classes 
(filled boxes), including the pp reference and normalization, is displayed 
separately from the remaining uncorrelated systematic uncertainty 
(open boxes). The $R_{\rm AA}$ of muons from heavy flavor decays at forward 
rapidity exhibits a strong suppression with increasing centrality, reaching a 
factor of about 3--4 in the 10$\%$ most central collisions.

The ALICE Collaboration has measured the production of prompt $\rm D$ mesons 
in $ 2 < p_{\rm t} < 16$~GeV/c at mid-rapidity 
($\vert y \vert < 0.5$)~\cite{Dmes12} and the CMS Collaboration reported on 
that of non-prompt $\rm J/\psi$ from beauty decays, in 
$6.5 < p_{\rm t} < 30$~\GeV\ and $\vert y \vert < 2.4$~\cite{cms}. 
The corresponding suppression of $\rm D$ mesons and $\rm J/\psi$ from beauty 
decays in those studies is similar to that reported here for muons from heavy 
flavor decays, although in a different \pt\ and rapidity region.

In conclusion, we have reported on the first measurement of the production of 
high-\pt\ muons from heavy flavor decays at forward rapidity, in pp and 
Pb--Pb collisions at $\sqrt {s_{\rm NN}}$~=~2.76 TeV with the ALICE detector. 
FONLL pQCD calculations describe well the pp data within experimental and 
theoretical uncertainties, with the data being close to the upper limit of 
the model predictions. The \RAA\ of high \pt\ muons from heavy flavor decays 
indicates a clear suppression increasing towards the most central collisions. 
The measured suppression is almost independent of \pt, in the region 
$4 < p_{\rm t} < 10$~\GeV. These results provide clear evidence for 
large in-medium effects for heavy quarks in central Pb--Pb collisions at 
$\sqrt {s_{\rm NN}}$~=~2.76 TeV. The forthcoming p--Pb collisions will 
complement these measurements, by providing insight into the possible 
contribution of initial nuclear matter effects, although those are 
expected to be less important in the high \pt\ region studied here

%
\newenvironment{acknowledgement}{\relax}{\relax}
\begin{acknowledgement}
\section{Acknowledgements}
The ALICE collaboration would like to thank all its engineers and technicians for their invaluable contributions to the construction of the experiment and the CERN accelerator teams for the outstanding performance of the LHC complex.
\\
The ALICE collaboration acknowledges the following funding agencies for their support in building and
running the ALICE detector:
 \\
Calouste Gulbenkian Foundation from Lisbon and Swiss Fonds Kidagan, Armenia;
 \\
Conselho Nacional de Desenvolvimento Cient\'{\i}fico e Tecnol\'{o}gico (CNPq), Financiadora de Estudos e Projetos (FINEP),
Funda\c{c}\~{a}o de Amparo \`{a} Pesquisa do Estado de S\~{a}o Paulo (FAPESP);
 \\
National Natural Science Foundation of China (NSFC), the Chinese Ministry of Education (CMOE)
and the Ministry of Science and Technology of China (MSTC);
 \\
Ministry of Education and Youth of the Czech Republic;
 \\
Danish Natural Science Research Council, the Carlsberg Foundation and the Danish National Research Foundation;
 \\
The European Research Council under the European Community's Seventh Framework Programme;
 \\
Helsinki Institute of Physics and the Academy of Finland;
 \\
French CNRS-IN2P3, the `Region Pays de Loire', `Region Alsace', `Region Auvergne' and CEA, France;
 \\
German BMBF and the Helmholtz Association;
\\
General Secretariat for Research and Technology, Ministry of
Development, Greece;
\\
Hungarian OTKA and National Office for Research and Technology (NKTH);
 \\
Department of Atomic Energy and Department of Science and Technology of the Government of India;
 \\
Istituto Nazionale di Fisica Nucleare (INFN) of Italy;
 \\
MEXT Grant-in-Aid for Specially Promoted Research, Ja\-pan;
 \\
Joint Institute for Nuclear Research, Dubna;
 \\
National Research Foundation of Korea (NRF);
 \\
CONACYT, DGAPA, M\'{e}xico, ALFA-EC and the HELEN Program (High-Energy physics Latin-American--European Network);
 \\
Stichting voor Fundamenteel Onderzoek der Materie (FOM) and the Nederlandse Organisatie voor Wetenschappelijk Onderzoek (NWO), Netherlands;
 \\
Research Council of Norway (NFR);
 \\
Polish Ministry of Science and Higher Education;
 \\
National Authority for Scientific Research - NASR (Autoritatea Na\c{t}ional\u{a} pentru Cercetare \c{S}tiin\c{t}ific\u{a} - ANCS);
 \\
Federal Agency of Science of the Ministry of Education and Science of Russian Federation, International Science and
Technology Center, Russian Academy of Sciences, Russian Federal Agency of Atomic Energy, Russian Federal Agency for Science and Innovations and CERN-INTAS;
 \\
Ministry of Education of Slovakia;
 \\
Department of Science and Technology, South Africa;
 \\
CIEMAT, EELA, Ministerio de Educaci\'{o}n y Ciencia of Spain, Xunta de Galicia (Conseller\'{\i}a de Educaci\'{o}n),
CEA\-DEN, Cubaenerg\'{\i}a, Cuba, and IAEA (International Atomic Energy Agency);
 \\
Swedish Research Council (VR) and Knut $\&$ Alice Wallenberg
Foundation (KAW);
 \\
Ukraine Ministry of Education and Science;
 \\
United Kingdom Science and Technology Facilities Council (STFC);
 \\
The United States Department of Energy, the United States National
Science Foundation, the State of Texas, and the State of Ohio.
\end{acknowledgement}
\newpage
%
%
\appendix
\section{The ALICE Collaboration}
\label{app:collab}

\begingroup
\small
\begin{flushleft}
B.~Abelev\Irefn{org1234}\And
J.~Adam\Irefn{org1274}\And
D.~Adamov\'{a}\Irefn{org1283}\And
A.M.~Adare\Irefn{org1260}\And
M.M.~Aggarwal\Irefn{org1157}\And
G.~Aglieri~Rinella\Irefn{org1192}\And
A.G.~Agocs\Irefn{org1143}\And
A.~Agostinelli\Irefn{org1132}\And
S.~Aguilar~Salazar\Irefn{org1247}\And
Z.~Ahammed\Irefn{org1225}
% ...
\renewcommand\labelenumi{\textsuperscript{\theenumi}~}
\section*{Affiliation notes}
\renewcommand\theenumi{\roman{enumi}}
\begin{Authlist}
\item \Adef{M.V.Lomonosov Moscow State University, D.V.Skobeltsyn Institute of Nuclear Physics, Moscow, Russia}Also at: M.V.Lomonosov Moscow State University, D.V.Skobeltsyn Institute of Nuclear Physics, Moscow, Russia
\item \Adef{Institute of Nuclear Sciences, Belgrade, Serbia}Also at: "Vin\v{c}a" Institute of Nuclear Sciences, Belgrade, Serbia
\end{Authlist}
\section*{Collaboration Institutes}
\renewcommand\theenumi{\arabic{enumi}~}
\begin{Authlist}
\item \Idef{org1279}Benem\'{e}rita Universidad Aut\'{o}noma de Puebla, Puebla, Mexico
\item \Idef{org1220}Bogolyubov Institute for Theoretical Physics, Kiev, Ukraine
\item \Idef{org1262}Budker Institute for Nuclear Physics, Novosibirsk, Russia
\item \Idef{org1292}California Polytechnic State University, San Luis Obispo, California, United States
\item \Idef{org14939}Centre de Calcul de l'IN2P3, Villeurbanne, France
\item \Idef{org1197}Centro de Aplicaciones Tecnol\'{o}gicas y Desarrollo Nuclear (CEADEN), Havana, Cuba
\item \Idef{org1242}Centro de Investigaciones Energ\'{e}ticas Medioambientales y Tecnol\'{o}gicas (CIEMAT), Madrid, Spain
\item \Idef{org1244}Centro de Investigaci\'{o}n y de Estudios Avanzados (CINVESTAV), Mexico City and M\'{e}rida, Mexico
\item \Idef{org1335}Centro Fermi -- Centro Studi e Ricerche e Museo Storico della Fisica ``Enrico Fermi'', Rome, Italy
\item \Idef{org17347}Chicago State University, Chicago, United States
\item \Idef{org1288}Commissariat \`{a} l'Energie Atomique, IRFU, Saclay, France
\item \Idef{org1294}Departamento de F\'{\i}sica de Part\'{\i}culas and IGFAE, Universidad de Santiago de Compostela, Santiago de Compostela, Spain
\item \Idef{org1106}Department of Physics Aligarh Muslim University, Aligarh, India
\item \Idef{org1121}Department of Physics and Technology, University of Bergen, Bergen, Norway
\item \Idef{org1162}Department of Physics, Ohio State University, Columbus, Ohio, United States
\item \Idef{org1300}Department of Physics, Sejong University, Seoul, South Korea
\item \Idef{org1268}Department of Physics, University of Oslo, Oslo, Norway
\item \Idef{org1145}Dipartimento di Fisica dell'Universit\`{a} and Sezione INFN, Cagliari, Italy
\item \Idef{org1270}Dipartimento di Fisica dell'Universit\`{a} and Sezione INFN, Padova, Italy
\item \Idef{org1315}Dipartimento di Fisica dell'Universit\`{a} and Sezione INFN, Trieste, Italy
\item \Idef{org1132}Dipartimento di Fisica dell'Universit\`{a} and Sezione INFN, Bologna, Italy
\item \Idef{org1285}Dipartimento di Fisica dell'Universit\`{a} `La Sapienza' and Sezione INFN, Rome, Italy
\item \Idef{org1154}Dipartimento di Fisica e Astronomia dell'Universit\`{a} and Sezione INFN, Catania, Italy
\item \Idef{org1290}Dipartimento di Fisica `E.R.~Caianiello' dell'Universit\`{a} and Gruppo Collegato INFN, Salerno, Italy
\item \Idef{org1312}Dipartimento di Fisica Sperimentale dell'Universit\`{a} and Sezione INFN, Turin, Italy
\item \Idef{org1103}Dipartimento di Scienze e Innovazione Tecnologica dell'Universit\`{a} del Piemonte Orientale and Gruppo Collegato INFN, Alessandria, Italy
\item \Idef{org1114}Dipartimento Interateneo di Fisica `M.~Merlin' and Sezione INFN, Bari, Italy
\item \Idef{org1237}Division of Experimental High Energy Physics, University of Lund, Lund, Sweden
\item \Idef{org1192}European Organization for Nuclear Research (CERN), Geneva, Switzerland
\item \Idef{org1227}Fachhochschule K\"{o}ln, K\"{o}ln, Germany
\item \Idef{org1122}Faculty of Engineering, Bergen University College, Bergen, Norway
\item \Idef{org1136}Faculty of Mathematics, Physics and Informatics, Comenius University, Bratislava, Slovakia
\item \Idef{org1274}Faculty of Nuclear Sciences and Physical Engineering, Czech Technical University in Prague, Prague, Czech Republic
\item \Idef{org1229}Faculty of Science, P.J.~\v{S}af\'{a}rik University, Ko\v{s}ice, Slovakia
\item \Idef{org1184}Frankfurt Institute for Advanced Studies, Johann Wolfgang Goethe-Universit\"{a}t Frankfurt, Frankfurt, Germany
\item \Idef{org1215}Gangneung-Wonju National University, Gangneung, South Korea
\item \Idef{org1212}Helsinki Institute of Physics (HIP) and University of Jyv\"{a}skyl\"{a}, Jyv\"{a}skyl\"{a}, Finland
\item \Idef{org1203}Hiroshima University, Hiroshima, Japan
\item \Idef{org1329}Hua-Zhong Normal University, Wuhan, China
\item \Idef{org1254}Indian Institute of Technology, Mumbai, India
\item \Idef{org36378}Indian Institute of Technology Indore (IIT), Indore, India
\item \Idef{org1266}Institut de Physique Nucl\'{e}aire d'Orsay (IPNO), Universit\'{e} Paris-Sud, CNRS-IN2P3, Orsay, France
\item \Idef{org1277}Institute for High Energy Physics, Protvino, Russia
\item \Idef{org1249}Institute for Nuclear Research, Academy of Sciences, Moscow, Russia
\item \Idef{org1320}Nikhef, National Institute for Subatomic Physics and Institute for Subatomic Physics of Utrecht University, Utrecht, Netherlands
\item \Idef{org1250}Institute for Theoretical and Experimental Physics, Moscow, Russia
\item \Idef{org1230}Institute of Experimental Physics, Slovak Academy of Sciences, Ko\v{s}ice, Slovakia
\item \Idef{org1127}Institute of Physics, Bhubaneswar, India
\item \Idef{org1275}Institute of Physics, Academy of Sciences of the Czech Republic, Prague, Czech Republic
\item \Idef{org1139}Institute of Space Sciences (ISS), Bucharest, Romania
\item \Idef{org27399}Institut f\"{u}r Informatik, Johann Wolfgang Goethe-Universit\"{a}t Frankfurt, Frankfurt, Germany
\item \Idef{org1185}Institut f\"{u}r Kernphysik, Johann Wolfgang Goethe-Universit\"{a}t Frankfurt, Frankfurt, Germany
\item \Idef{org1177}Institut f\"{u}r Kernphysik, Technische Universit\"{a}t Darmstadt, Darmstadt, Germany
\item \Idef{org1256}Institut f\"{u}r Kernphysik, Westf\"{a}lische Wilhelms-Universit\"{a}t M\"{u}nster, M\"{u}nster, Germany
\item \Idef{org1246}Instituto de Ciencias Nucleares, Universidad Nacional Aut\'{o}noma de M\'{e}xico, Mexico City, Mexico
\item \Idef{org1247}Instituto de F\'{\i}sica, Universidad Nacional Aut\'{o}noma de M\'{e}xico, Mexico City, Mexico
\item \Idef{org23333}Institut of Theoretical Physics, University of Wroclaw
\item \Idef{org1308}Institut Pluridisciplinaire Hubert Curien (IPHC), Universit\'{e} de Strasbourg, CNRS-IN2P3, Strasbourg, France
\item \Idef{org1182}Joint Institute for Nuclear Research (JINR), Dubna, Russia
\item \Idef{org1143}KFKI Research Institute for Particle and Nuclear Physics, Hungarian Academy of Sciences, Budapest, Hungary
\item \Idef{org1199}Kirchhoff-Institut f\"{u}r Physik, Ruprecht-Karls-Universit\"{a}t Heidelberg, Heidelberg, Germany
\item \Idef{org20954}Korea Institute of Science and Technology Information, Daejeon, South Korea
\item \Idef{org1160}Laboratoire de Physique Corpusculaire (LPC), Clermont Universit\'{e}, Universit\'{e} Blaise Pascal, CNRS--IN2P3, Clermont-Ferrand, France
\item \Idef{org1194}Laboratoire de Physique Subatomique et de Cosmologie (LPSC), Universit\'{e} Joseph Fourier, CNRS-IN2P3, Institut Polytechnique de Grenoble, Grenoble, France
\item \Idef{org1187}Laboratori Nazionali di Frascati, INFN, Frascati, Italy
\item \Idef{org1232}Laboratori Nazionali di Legnaro, INFN, Legnaro, Italy
\item \Idef{org1125}Lawrence Berkeley National Laboratory, Berkeley, California, United States
\item \Idef{org1234}Lawrence Livermore National Laboratory, Livermore, California, United States
\item \Idef{org1251}Moscow Engineering Physics Institute, Moscow, Russia
\item \Idef{org1140}National Institute for Physics and Nuclear Engineering, Bucharest, Romania
\item \Idef{org1165}Niels Bohr Institute, University of Copenhagen, Copenhagen, Denmark
\item \Idef{org1109}Nikhef, National Institute for Subatomic Physics, Amsterdam, Netherlands
\item \Idef{org1283}Nuclear Physics Institute, Academy of Sciences of the Czech Republic, \v{R}e\v{z} u Prahy, Czech Republic
\item \Idef{org1264}Oak Ridge National Laboratory, Oak Ridge, Tennessee, United States
\item \Idef{org1189}Petersburg Nuclear Physics Institute, Gatchina, Russia
\item \Idef{org1170}Physics Department, Creighton University, Omaha, Nebraska, United States
\item \Idef{org1157}Physics Department, Panjab University, Chandigarh, India
\item \Idef{org1112}Physics Department, University of Athens, Athens, Greece
\item \Idef{org1152}Physics Department, University of Cape Town, iThemba LABS, Cape Town, South Africa
\item \Idef{org1209}Physics Department, University of Jammu, Jammu, India
\item \Idef{org1207}Physics Department, University of Rajasthan, Jaipur, India
\item \Idef{org1200}Physikalisches Institut, Ruprecht-Karls-Universit\"{a}t Heidelberg, Heidelberg, Germany
\item \Idef{org1325}Purdue University, West Lafayette, Indiana, United States
\item \Idef{org1281}Pusan National University, Pusan, South Korea
\item \Idef{org1176}Research Division and ExtreMe Matter Institute EMMI, GSI Helmholtzzentrum f\"ur Schwerionenforschung, Darmstadt, Germany
\item \Idef{org1334}Rudjer Bo\v{s}kovi\'{c} Institute, Zagreb, Croatia
\item \Idef{org1298}Russian Federal Nuclear Center (VNIIEF), Sarov, Russia
\item \Idef{org1252}Russian Research Centre Kurchatov Institute, Moscow, Russia
\item \Idef{org1224}Saha Institute of Nuclear Physics, Kolkata, India
\item \Idef{org1130}School of Physics and Astronomy, University of Birmingham, Birmingham, United Kingdom
\item \Idef{org1338}Secci\'{o}n F\'{\i}sica, Departamento de Ciencias, Pontificia Universidad Cat\'{o}lica del Per\'{u}, Lima, Peru
\item \Idef{org1316}Sezione INFN, Trieste, Italy
\item \Idef{org1271}Sezione INFN, Padova, Italy
\item \Idef{org1313}Sezione INFN, Turin, Italy
\item \Idef{org1286}Sezione INFN, Rome, Italy
\item \Idef{org1146}Sezione INFN, Cagliari, Italy
\item \Idef{org1133}Sezione INFN, Bologna, Italy
\item \Idef{org1115}Sezione INFN, Bari, Italy
\item \Idef{org1155}Sezione INFN, Catania, Italy
\item \Idef{org1322}Soltan Institute for Nuclear Studies, Warsaw, Poland
\item \Idef{org36377}Nuclear Physics Group, STFC Daresbury Laboratory, Daresbury, United Kingdom
\item \Idef{org1258}SUBATECH, Ecole des Mines de Nantes, Universit\'{e} de Nantes, CNRS-IN2P3, Nantes, France
\item \Idef{org1304}Technical University of Split FESB, Split, Croatia
\item \Idef{org1168}The Henryk Niewodniczanski Institute of Nuclear Physics, Polish Academy of Sciences, Cracow, Poland
\item \Idef{org17361}The University of Texas at Austin, Physics Department, Austin, TX, United States
\item \Idef{org1173}Universidad Aut\'{o}noma de Sinaloa, Culiac\'{a}n, Mexico
\item \Idef{org1296}Universidade de S\~{a}o Paulo (USP), S\~{a}o Paulo, Brazil
\item \Idef{org1149}Universidade Estadual de Campinas (UNICAMP), Campinas, Brazil
\item \Idef{org1239}Universit\'{e} de Lyon, Universit\'{e} Lyon 1, CNRS/IN2P3, IPN-Lyon, Villeurbanne, France
\item \Idef{org1205}University of Houston, Houston, Texas, United States
\item \Idef{org20371}University of Technology and Austrian Academy of Sciences, Vienna, Austria
\item \Idef{org1222}University of Tennessee, Knoxville, Tennessee, United States
\item \Idef{org1310}University of Tokyo, Tokyo, Japan
\item \Idef{org1318}University of Tsukuba, Tsukuba, Japan
\item \Idef{org21360}Eberhard Karls Universit\"{a}t T\"{u}bingen, T\"{u}bingen, Germany
\item \Idef{org1225}Variable Energy Cyclotron Centre, Kolkata, India
\item \Idef{org1306}V.~Fock Institute for Physics, St. Petersburg State University, St. Petersburg, Russia
\item \Idef{org1323}Warsaw University of Technology, Warsaw, Poland
\item \Idef{org1179}Wayne State University, Detroit, Michigan, United States
\item \Idef{org1260}Yale University, New Haven, Connecticut, United States
\item \Idef{org1332}Yerevan Physics Institute, Yerevan, Armenia
\item \Idef{org15649}Yildiz Technical University, Istanbul, Turkey
\item \Idef{org1301}Yonsei University, Seoul, South Korea
\item \Idef{org1327}Zentrum f\"{u}r Technologietransfer und Telekommunikation (ZTT), Fachhochschule Worms, Worms, Germany
\end{Authlist}
\endgroup

%
%

\begin{thebibliography}{99}
\bibitem{note} In the ALICE reference frame, the muon spectrometer covers a 
negative $\eta$ range and consequently a negative $y$ range. The results are 
presented with a positive $y$ notation.
\end{thebibliography}
\end{document}